\documentclass[12pt]{article}
\usepackage{psfig,aaspp4,flushrt}

\lefthead{Lehnert, Heckman, \& Weaver}
\righthead{Very Extended Emission in M~82}

\begin{document}

\title{Very Extended X-ray and H$\alpha$ Emission in M~82: Implications
for the Superwind Phenomenon}
\author{Matthew D. Lehnert\altaffilmark{1}}
\affil{Max-Plank-Institut f\"ur extraterrestrische Physik, Postfach 1603, D-85740 Garching, Germany}
\medskip
\medskip
\author{Timothy M. Heckman}
\affil{Department of Physics and Astronomy,
Johns Hopkins University, Baltimore, MD 21218}
\medskip
\and
\medskip
\author{Kimberly A. Weaver}
\affil{NASA Goddard Space Flight
Center, Greenbelt, MD 20771 and Department of Physics and Astronomy,
Johns Hopkins University, Baltimore, MD 21218}

\medskip

\altaffiltext{1}{Visiting observer at Kitt Peak National Observatory of
the National Optical Astronomy Observatories, operated by AURA under
contract with the National Science Foundation.}

\medskip



\newpage

\begin{abstract}

We discuss the properties and implications of the 3.5 $\times$ 0.9 arc
minute (3.7 $\times$ 0.9 kpc) region of spatially-coincident X-ray and
H$\alpha$ emission about 11 arc minutes (11.6 kpc) to the north of the
prototypical starburst/superwind galaxy M~82 previously discussed by
\cite{DB99}.  The total H$\alpha$ flux from this ridge of
emission is 1.5 $\times$ 10$^{-13}$ ergs s$^{-1}$ cm $^{-2}$, or about
0.3\% of the total M~82 H$\alpha$ flux. The implied H$\alpha$ luminosity
of this region is 2.4 $\times$ 10$^{38}$ ergs s$^{-1}$.  Diffuse soft
X-ray emission is seen over the same region by the ROSAT PSPC and HRI.
The PSPC X-ray spectrum is fit by thermal plasma absorbed by only the
Galactic foreground column density (N$_H$ = 3.7 $\times$ 10$^{20}$
cm$^{-2}$) and having a temperature of kT=0.80$\pm$0.17 keV.  The total
unabsorbed flux from the ridge is 1.4 $\times$ 10$^{-13}$ ergs
cm$^{-2}$ s$^{-1}$ ($\sim$ 2.2 $\times$ 10$^{38}$ ergs s$^{-1}$),
comprising about 0.7\% of the total X-ray emission from M~82.

We evaluate the relationship of the X-ray/H$\alpha$ ridge to the M~82
superwind.  The H$\alpha$ emission could be excited by ionizing
radiation from the starburst that propagates into the galactic halo
along the cavity carved by the superwind. However, the main properties
of the X-ray emission can all be explained as being due to
shock-heating driven as the superwind encounters a massive ionized
cloud in the halo of M~82 (possibly related to the tidal debris seen in
HI in the interacting M~81/M~82/NGC~3077 system). This encounter drives a
slow shock into the cloud, which contributes to the excitation of the
observed H$\alpha$ emission. At the same time, a fast bow-shock
develops in the superwind just upstream of the cloud, and this produces
the observed X-ray emission. This interpretation would imply that the
superwind has an outflow speed of roughly 800 km s$^{-1}$, consistent
with indirect estimates based on its general X-ray properties and the
kinematics of the inner kpc-scale region of H$\alpha$ filaments. The
alternative, in which a much faster and more tenuous wind drives a fast
radiative shock into a cloud that then produces {\it both} the X-ray
and H$\alpha$ emission is ruled out by the long radiative cooling times
and the relatively quiescent H$\alpha$ kinematics in this region.

We suggest that wind-cloud interactions may be an important mechanism
for generating X-ray and optical line emission in the halos of
starbursts. Such interactions can establish that the wind has
propagated out to substantially greater radii than could otherwise be
surmised.  This has potentially interesting implications for the fate
of the outflowing metal enriched material, and bears on the role of
superwinds in the metal enrichment and heating of galactic halos and
the intergalactic medium.  In particular, the gas in the M~82 ridge is
roughly two orders-of-magnitude hotter than the minimum ``escape
temperature'' at this radius, so this gas will not be retained by M~82.
\end{abstract}

\keywords{Galaxies: active --- galaxies: nuclei ---
galaxies: starburst --- galaxies: ISM --- optical: galaxies ---
individual: M~82}

\newpage

\section{Introduction}

The history of the study of M~82 is a long and rich one.  M~82 is the
prototypical starburst galaxy and since it is one of the nearest (D = 3.63 Mpc
- \cite{F94}) and brightest members of that class, it has been studied
at almost every wavelength ranging from the $\gamma$-rays through to
low frequency radio waves (e.g., \cite{B94}; \cite{BST95};
\cite{C90};\cite{LS63}; \cite{SBH98}; \cite{L94}; \cite{P94};
\cite{SKB94}; \cite{H94}; \cite{R94}; \cite{ML97}; \cite{P97};
\cite{SPS97}).  It is perhaps impossible for any paper, let alone one
this brief, to do justice to this rich history.  We refer the reader to
excellent exposes on M~82 in, for example, \cite{BT88} and \cite{T88}.

M~82 is also the prototype of the ``superwind'' phenomenon
(\cite{AT78}; \cite{MHvB87}; \cite{BT88}; \cite{HAM90}; \cite{BST95};
\cite{SPS97}; \cite{SBH98}).  These flows are powered by the collective
kinetic energy and momentum injected by stellar winds and supernovae in
starburst galaxies.  While recent studies have established the ubiquity
of superwinds in local starburst galaxies (e.g., \cite{LH96};
\cite{DWH98}), there is still only a rough understanding of either
their dynamics or the processes responsible for the observed optical
and X-ray emission (e.g., \cite{Str98} and references therein). Without
such understanding, it is difficult to quantitatively assess the role
that superwinds have played in the formation and evolution of galaxies
(\cite{KC98}; \cite{SPF98}) and in the chemical enrichment,
heating, and possible magnetization of the inter-galactic medium
(\cite{GLM97}; \cite{PCN98}; \cite{KLH99}).

In this paper we discuss the optical and X-ray properties of the
spatially-correlated region of X-ray and H$\alpha$ line emission
recently reported by \cite{DB99} (DB hereafter) to be
located roughly 11 kpc above the disk of M~82, far beyond the region
where the superwind has been previously investigated.  We are motivated
by the important implications of understanding such a region for the
superwind theory.  If this region could be shown to be physically
related to the superwind that M~82 is driving, then it implies that
winds driven even by energetically-modest starbursts like M~82 are able
to reach large distances into the halo, making them more likely to be
able to escape the gravitational potential of the galaxy and thereby
supply mass, metal, and energy to the inter-galactic medium (IGM). Such
a finding would have implications for our understanding of the
evolution of the IGM, the halos of galaxies, and quasar absorption line
systems. Our study of this emission region also provides some unique
clues about the physics that underlies the observed X-ray and optical
emission from superwinds, and strongly suggests that wind-cloud
collisions in starburst galaxy halos are an important emission
process.

\section{Observations and Data Reduction}

\subsection{X-ray Data}

The ROSAT PSPC data were obtained during the period from March through
October 1991 and had a total integration time of about 25 ksec.  M~82
was placed approximately in the center of the field for all the
observations, and the source size is small enough so that there is no
obscuration of the X-ray emission from M~82 due to the mirror ribs.

We extracted a spectrum from a region of diffuse X-ray emission located
about 11 arc minutes north of M~82, and coincident with the region of
H$\alpha$ emission reported by DB.  We will subsequently refer to this
region as the ``ridge'' of emission.  This area of X-ray and optical
emission is oriented roughly parallel to the major axis of M~82, and was
denoted as sources 8 and 9 in \cite{DWH98}.  We extracted source counts
in an elliptical region with a major (minor) axis of 3.3 (1.8) arc
minutes oriented in a position angle of 43$^\circ$.  This region is just
sufficient to encompass all of the obvious X-ray emission, but excludes
the soft point source at the ENE end of the ridge of emission that is
discussed in \S 3.2 below.  We have also extracted source counts for the
portion of the ridge that is contaminated by this point source using a
circular aperture with a diameter of 1.75 arcmin.  The background for
these spectra was taken from a large region where there were no apparent
background sources to the south and west of M~82.  The relative distance
of the background region from the center of the PSPC field was
approximately the same as that for the ridge of emission.  The spectrum
was then grouped such that each channel contained at least 25 counts.
The extracted background subtracted data were subsequently fitted with
various models using the package XSPEC.

We have also inspected the archival ROSAT HRI data for M~82. These data
have lower sensitivity than the PSPC data, and provide no spectral
information. However, they do provide significantly higher spatial
resolution ($\sim$6 arcsec FWHM on-axis and $\sim$15 arcsec at the
position of the X-ray ridge).  The data we have used comprise a 53 ksec
exposure taken in April and May 1995. We have used these data only to
study the morphology of the X-ray ridge.

\subsection{Optical Imaging}

Our optical images of M~82 were taken on the night of February 27, 1995
(UT) using the KPNO 0.9m telescope.  We observed with a Tek 2048$^2$
CCD and through the optics of 0.9m, this yielded a scale of 0.68 arc
seconds pixel$^{-1}$ and a total field of view of over 23 arc minutes.
The night was not photometric, with cirrus around all night and
intermittent patches of thicker clouds.  M~82 was observed through two
filters, a narrow-band filter with a full width at half maximum of
29\AA \ and a central wavelength of 6562\AA \ and a wider filter with a
FWHM of 85\AA \ and a central wavelength of 6658\AA.  The first filter
effectively isolated H$\alpha$ and underlying continuum and did not
include emission from either [NII] lines at 6548 and 6584\AA.  The
second filter provided a line-free measure of the continuum and was
used for continuum subtraction of the first narrow-band filter to
produce a H$\alpha$ line-only image of M~82.  The total integration time
in each filter was 1200 seconds.

The images were reduced in the usual way (bias subtracted, flat-fielded,
and then the narrow-band continuum image was aligned and scaled to
subtract the continuum from the narrow-band H$\alpha$ image) using the
reduction package IRAF\footnote{2}{IRAF is distributed by the National
Optical Astronomy Observatories, which are operated by the Association
of Universities for Research in Astronomy, Inc., under cooperative
agreement with the National Science Foundation}.

Since the data were taken under non-photometric conditions, fluxing the
data directly using observations of spectrophotometric standards is
impossible.  However, obtaining fluxes and hence luminosities is very
important to our analysis.  Therefore, to estimate the H$\alpha$ flux
and continuum flux-density of the off-band filter of the continuum and
emission-line nebulae of M~82, we bootstrapped from the flux given in a
5.8 arc second circular aperture centered on knot C from \cite{OM78}.
Knot C was chosen because it was easily identifiable in our images and
it appears in a relatively uncomplicated region of line emission.  This
procedure yields a H$\alpha$ flux from M~82 of about 4.6 $\times$
10$^{-11}$ ergs cm$^{-2}$ s$^{-1}$, which compares very well with the
H$\alpha$ flux from M~82 estimated by \cite{MHvB87} of
4.5$\times$10$^{-11}$ ergs cm$^{-2}$ s$^{-1}$ over a similar region 90"
$\times$ 90".  All subsequent results for the H$\alpha$ flux and
luminosity of various features within the H$\alpha$ image of M~82 will
rely on this bootstrapping procedure.

To estimate the continuum flux-density we used the flux-density of
O'Connell \& Mangano taken at 6560\AA.  Since the center of our
narrow-band continuum emission is 6658\AA, we are thus assuming that the
continuum has a constant flux-density between 6560\AA \ and 6658\AA.  To
see if this extrapolation procedure is accurate, we compared the flux in
R of M~82 in a 35 arc second aperture of 1.97$\times$ 10$^{-13}$ ergs
cm$^{-2}$ s$^{-1}$\AA$^{-1}$ (from NED based on data originally obtained
by \cite{J66}).  If we also take the total counts in the 6658\AA \
filter and determine the flux-density by extrapolating the conversion
factor from measuring knot C of O'Connell \& Mangano, we find a
flux-density of 2.11 $\times$ 10$^{-13}$ ergs cm$^{-2}$ s$^{-1}$
\AA$^{-1}$ in a 35 arc second aperture.  Our extrapolated flux is within
10\% (0.1 magnitudes) and thus this extrapolation is reasonably accurate.

\section{Results}

\subsection{Optical Imaging}

The H$\alpha$ and narrow-band continuum images of M~82 are quite
spectacular (Figures 1 and 2).  We find that the line emission is very
extended, with discernible H$\alpha$ line emission extending to about
11 kpc above the plane of M~82. This very faint emission can be seen
even more clearly in the deeper H$\alpha$ image obtained by DB.  Like
previous authors (e.g., \cite{BT88}; \cite{MHvB87}), we
find that the H$\alpha$ line emission has a very complex morphology
that is very suggestive of outflowing gas, with loops, tendrils, and
filaments of line emission pointing out of the plane of the galaxy.

An especially interesting feature reported by DB is the faint ridge of
emission about 11 arc minutes north of M~82 at a position angle of about
$-$20$^\circ$ from the nucleus.  This ridge is then about 11.6 kpc (in
projection) north of M~82 at the adopted distance of 3.63 Mpc
(\cite{F94}).  The dimensions of the ridge down to surface brightnesses
of about 3.5 $\times$ 10$^{-17}$ ergs s$^{-1}$ cm $^{-2}$ arcsec$^{-2}$
are about 3.5 arc minutes long $\times$ 0.9 arc minutes wide, with the
long axis being approximately parallel to the major axis of M~82.  At
the assumed distance of M~82, this size translates into approximately
3.7 kpc $\times$ 0.9 kpc and the total H$\alpha$ flux from this ridge
of emission is 1.5 $\times$ 10$^{-13}$ ergs s$^{-1}$ cm $^{-2}$.  This
flux is reasonably close to the estimate in DB of 1.2 $\times$
10$^{-13}$ ergs s$^{-1}$ cm$^{-2}$, and yields an H$\alpha$ luminosity
of 2.4$\times$ 10$^{38}$ ergs s$^{-1}$.  The total H$\alpha$ flux for
the whole of M~82 is approximately 4.6 $\times$ 10$^{-11}$ ergs s$^{-1}$
cm $^{-2}$. Thus, the ridge of emission constitutes approximately 0.3\%
of the total H$\alpha$ emission from M~82 (not corrected for internal
extinction).  While the ridge appears rather structureless in our
H$\alpha$ image, the deeper image published by DB shows a collection of
bright knots and loops or filaments connected to and immersed in the
more diffuse and pervasive emission.

The narrow-band line-free {\it continuum} image reveals a complex
morphology that is as interesting as the H$\alpha$ image.  Besides the
obvious complex dust absorption that is seen in the image (with some
dark features oriented perpendicular to the plane of the galaxy), we
also see filaments of continuum emission extended out of the plane of
the galaxy (although the filaments are not entirely obvious in Figure
2).  These filaments have twists and turns as seen projected on the sky,
but generally point perpendicular to the plane of the galaxy.  They
extend outwards almost 3 arc minutes to the northwest and almost 2.5 arc
minutes to the southeast -- corresponding to a projected distance out of
the plane of about 3 kpc.  Even though the continuum emission is in many
ways as complicated as that seen in the H$\alpha$ image, we do not
observe any continuum emission from the ridge of spatially coincident
H$\alpha$ -- X-ray emission.  Using the counts to flux-density
conversion for the narrow-band line-free continuum image and the same
aperture as used to measure the H$\alpha$ flux (but with the point
sources removed from the area), we find an upper limit for the possible
continuum emission from the ridge of fainter than 15.8 R magnitudes.
This then implies an absolute R magnitude fainter than $-$12.0.

While we find no evidence for any diffuse, spatially-extended optical
continuum associated with the H$\alpha$ ridge, we do find a relatively
bright point source in the line-free image that is coincident with the
X-ray point source seen at the ENE end of the ridge in the ROSAT HRI
data. This point source has an R magnitude of 12.2. We will discuss its
likely nature in the next section.

\subsection{X-ray Imaging and Spectroscopy}

The optical images described above were aligned with the X-ray images
starting with the coordinates given for the X-ray data from ROSAT and
using the positions of stars within the frame from the Guide Star
Catalogue and using coordinates given in the header from the digital
sky survey.  We then slightly adjusted the X-ray pointing by
identifying and aligning X-ray sources on the digital sky survey image
of the region. The close spatial correspondence between the ridge of
optical and X-ray emission about 11 arcmin to the North of M~82 is
obvious (Figure 1).

We do not display the ROSAT HRI image of the ridge, since these data
show only that the ridge is largely diffuse and structureless with the
15 arcsec (260 pc) off-axis spatial resolution of the HRI. The only
substructure is a point source at the easternmost end of the ridge.
This source appears to coincide with a stellar object visible in Figure
2 (the relatively bright star-like object located about 40 arcsec ENE
of knot `I' in Figure 2 of DB). This is most likely a foreground star,
as we will briefly discuss below.

Examination of the 0.25 keV, 0.75 keV, and 1.5 keV PSPC maps of M~82
presented by \cite{DWH98} -- hereafter DWH -- shows
that the X-ray morphology of the ridge is energy-dependent. The ENE
portion of the ridge (source 9 in DWH) is relatively much more
conspicuous in the 0.25 keV map compared to the rest of the ridge
(source 8 in DWH). This reflects the contamination of the eastern-most
ridge emission by the (presumably) unrelated point-source visible in
the HRI image.

As described in \S 2.1 above, we have therefore extracted a PSPC
spectrum of the main body of the ridge from a region that excluded the
point source.  The relatively small number of detected photons precludes
fitting complex multi-component models, and we fit a model of a single
thermal (``MEKAL'') component to the spectrum.  We are motivated to use
a thermal model by the detailed X-ray spectroscopy of the brighter
portions of the M~82 outflow (DWH; \cite{ML97}; \cite{SPS97}).  The fits
strongly preferred a low absorbing column, so we simply froze the column
at the Galactic value of N$_H$ = 3.7 $\times$ 10$^{20}$ cm$^{-2}$.  The
data quality were not adequate to fit for the metal abundance, so we
froze this at the solar value.  Changing the assumed metal abundance did
not significantly affect the best-fit temperature, but does directly
affect the normalization of the fit.  The best-fit temperature was
kT=0.76$\pm$0.13 keV (90\% confidence interval with a reduced $\chi^{2}$
$\approx$ 1.3; Figure 3).  The relatively high temperature is similar to
that measured in the ROSAT PSPC spectroscopy of the innermost region of
the M~82 outflow ($\sim$0.7 keV; DWH; \cite{SPS97}).  The total
unabsorbed flux in the 0.1 -- 2.4 keV band from this part of the ridge
of emission is 8.5 $\times$ 10$^{-14}$ ergs cm$^{-2}$ s$^{-1}$.

For the ENE part of the ridge - whose emission is contaminated by the
point source described above - we have fit a two-component MEKAL model
(Figure 3).  Both components were frozen at solar abundances, one
component (the ridge emission) had an absorbing column frozen at N$_H$ =
3.7 $\times$ 10$^{20}$ cm$^{-2}$, and the other (the stellar point
source) had an absorbing column frozen at 0.  This fit yields
kT=0.85$\pm$0.17 keV for the ridge and kT$\sim$0.1 keV for the point
source.  The corresponding unabsorbed fluxes are 5.8 $\times$ 10$^{-14}$
ergs cm$^{-2}$ s$^{-1}$ and 3.5 $\times$ 10$^{-15}$ ergs cm$^{-2}$
s$^{-1}$ for the ridge and point source respectively.  Combining the
fits for the two regions of the ridge, the total X-ray flux is then 1.4
$\times$ 10$^{-13}$ ergs cm$^{-2}$ s$^{-1}$.  This comprises about 0.7\%
of the total unabsorbed X-ray emission from M~82 in the PSPC band (e.g.,
\cite{ML97}; DWH).  The total X-ray luminosity of the ridge of emission
is 2.2$\times$10$^{38}$ ergs s$^{-1}$ (in the 0.1 -- 2.4 keV band).

As stated above, the ENE point source is identified with a stellar
object having an R magnitude of 12.2 ($\lambda$F$_{\lambda}$ = 1.6
$\times$ 10$^{-10}$ erg cm$^{-2}$ s$^{-1}$).  The X-ray/optical flux
ratio suggests that the most likely nature of this source is a
late-type dwarf star, and this would also be consistent with its soft
ROSAT PSPC spectrum (e.g., \cite{RGV85}).

\section{Discussion}

\subsection{Is the Ridge of Emission Associated with M~82?}

Being able to isolate the ridge of emission to the north of M~82 allows
us to investigate its nature separately from the rest of M~82.  Could
this ridge of emission be due to another galaxy near M~82, or is it
instead gas excited by the M~82 starburst?

We have found that the ridge of X-ray and H$\alpha$ emission has fairly
high X-ray and H$\alpha$ luminosities -- both luminosities are in the
range observed for star-forming dwarf galaxies (e.g., \cite{HHG93};
\cite{H95}).  However, we see no evidence for any optical continuum
emission from this ridge. The limit for the level of continuum emission
implies that any putative galaxy must be very low luminosity,
comparable to that of the local dwarf spheroidal galaxies
(\cite{M98}).  Moreover, the limit on the continuum emission implies an
H$\alpha$ equivalent width of greater than 180\AA.  Examining the
sample of galaxies with M$_B$ fainter than $-$15 in the the H$\alpha$
survey of nearby dwarf irregular galaxies by \cite{HHG93}, we find that
these have typical global H$\alpha$ equivalent widths of a few to a few
tens of \AA.  The only such galaxy with an H$\alpha$ equivalent width
comparable to our lower limit for the M~82 ridge is DDO53. This galaxy
has a ratio of HI mass to H$\alpha$ luminosity (in solar units) of 420.
The corresponding HI mass for the M~82 ridge would have to be 2.5
$\times 10^7$ M$_{\odot}$. The HI map in \cite{Y93} implies a maximum
HI mass in the M~82 ridge that is only about 10$^6$ M$_{\sun}$.
Moreover, while the M~82 ridge has an X-ray luminosity similar to that
of the dwarf starburst galaxy NGC1569 (\cite{H95}), the M~82 ridge would
have a ratio of X-ray to optical continuum luminosity that would be
over 40 times higher than NGC1569.  Thus, it is clear that the ridge in
M~82 has properties that are very different from those of star-forming
dwarf galaxies. This argument, together with the strong morphological
connection between the M~82 ridge and the M~82 outflow (see DB), leads us
to conclude that this ridge is {\it not} a dwarf star-forming galaxy.

Other possibilities for this region, like a cloud in the halo of the
Milky Way are also highly unlikely.  The HI detected in the data of
\cite{Y93} and the H$\alpha$ emission from the spectrum of DB have
velocities such that it they are both likely to be associated with M~82
and not the galaxy (\cite{BH97}; \cite{hrt98}; \cite{r95}).  Moreover,
while galactic high velocity HI clouds often show soft X-ray excesses
(\cite{Kerp96}; \cite{Kerp99}), the associated X-ray emission is very
soft ($\approx$0.1-0.2 keV; see discussion in \cite{Kerp96}).

\subsection{Photoionization of the Ridge by M~82}

The ridge is evidently related to, and excited by M~82.  In the case of
the H$\alpha$ emission two possibilities are obvious.  Either the
superwind in M~82 has worked its way out to $\approx$11 kpc above the
plane of the galaxy and is shocking halo material that appears in
plentiful supply around M~82 (\cite{Y93}, 1994), or Lyman continuum
radiation  from the starburst is ionizing gas at very large distances
from the disk. Only the first possibility could apply to the X-ray
emission from the ridge.

Let us suppose that ambient gas is being photoionized by the stellar
population of M~82.  The rate of ionizing photons from M~82 is
approximately $10^{54}$ s$^{-1}$ (\cite{McL93} and references
therein).  Using the projected size and distance from M~82 estimated for
the ridge of emission given in the previous section, and assuming that
the ridge is really a ``box'' with a depth equal to its projected
length (i.e., 3.7 kpc $\times$ 3.7 kpc) we find that the ridge will
intercept about 8 $\times$ 10$^{51}$ ionizing photons s$^{-1}$.  This
of course is an upper limit since some (most?) of the ionizing photons
will be absorbed or dust-scattered before they reach distances as great
as 11.6 kpc out of the plane of M~82.  We can estimate the rate of
photons necessary to give the H$\alpha$ luminosity of the ridge (2.4
$\times$ 10$^{38}$ ergs s$^{-1}$) by assuming that the gas is in
photo-ionization equilibrium, i.e., the number of recombinations equals
the number of absorptions.  Using the numbers for the Case B effective
recombination rate of H$\alpha$ from \cite{O89} we find that it
is necessary to have a rate of $\sim$ 2$\times$ 10$^{50}$ ionizing
photons s$^{-1}$.  This is $\sim$40 times smaller than the upper limit
for the number of photons that M~82 is able to provide, so only about 2
to 3\% of the ionizing radiation initially emitted in the direction of
the ridge needs to make it to this distance. Such an `escape fraction'
is consistent with HUT spectra of starbursts below the Lyman edge
(\cite{L95}; \cite{Hur97}), and it seems plausible that
the optimal path along which ionizing radiation could escape from a
starburst is along the cavity carved-out by the superwind.

DB have made similar arguments in favor of a starburst-photoionization
origin for the H$\alpha$ emission.  However, photoionization by stars
does not produce the hot X-ray emission that is coincident with the
H$\alpha$ emission. Instead, shock-heating by the superwind may be an
efficient mechanism for producing coincident X-ray and line-emitting
plasma. We now explore the energetic plausibility of this model, and
revisit the issue of the role of photoionization in \S 4.5.

\subsection{Shock Energetics and the Basic Physical Properties of the Ridge}

The ridge of emission has an X-ray luminosity of 2.2 $\times$ 10$^{38}$
ergs s$^{-1}$ and a temperature of 9$\times$ 10$^6$ K.  Assuming that
the shock is strong and adiabatic, we have that $T_{post-shock}={3
\over 16} {\mu v^2_s \over k}$, where $\mu$ is the mass per particle,
$v_s$ is the shock velocity, and $k$ is Boltzmann's constant.  Thus to
produce gas with T = 9 $\times$ 10$^6$ K requires a shock speed of 820
km s$^{-1}$.  Such a high shock speed is quite reasonable considering
that M~82 has been estimated to have a deprojected outflow velocity in
optical line emitting filaments observed along the minor axis of
roughly 660 km s$^{-1}$ (\cite{SBH98}), and the outflowing X-ray gas
could {\it in principle} have a velocity as high as 3000 km s$^{-1}$
(\cite{CC85}). It would take material traveling at 820 km s$^{-1}$,
$\approx$1.4 $\times$ 10$^7$ yrs to travel the 11.6 kpc from M~82 to the
ridge of emission.  Since this is approximately the estimated lifetime
of the starburst in M~82 (\cite{R93}; \cite{B92}; \cite{Sch98};
\cite{S97}), it is very plausible that the starburst could drive a shock
over its lifetime this far out of the galactic plane.

If we assume that we are observing a ``box'' of emission with
dimensions of 3.7 kpc $\times$ 3.7 kpc $\times$ 0.9 kpc, the total
volume is 3.7 $\times$ 10$^{65}$ cm$^{-3}$.  Assuming this region is
occupied by hot gas filling a fraction f$_X$ of this volume, the
measured luminosity in the ROSAT band pass of 0.1 to 2.4 keV implies
that the X-ray emitting gas has an electron density, n$_{e,X}$ = 5.4
$\times$ 10$^{-3}$ f$_X$$^{-1/2}$ cm$^{-3}$, pressure, P$_X\approx$
2n$_{e,X}$kT=1.3 $\times$ 10$^{-11}$ f$_X$$^{-1/2}$ dynes cm$^{-2}$,
cooling time, t$_{cool} \approx$ $3kT/n_{e,X}\Lambda$=2.7 $\times$
10$^8$ f$_X$$^{1/2}$ yrs, a mass, M$_{X}$=2.0 $\times$ 10$^6$
f$_X$$^{1/2}$ M$_{\sun}$, and a total thermal energy of 8$\times$
10$^{54}$ f$_X$$^{1/2}$ ergs. Similarly, assuming that conditions of
Case B recombination and a temperature of 10$^4$ K holds in the region
producing the H$\alpha$ emission, and that this region also occupies a
volume 3.7 $\times$ 10$^{65}$ f$_{H\alpha}$ cm$^{-3}$, then the implied
density and mass of this gas is n$_{e,H\alpha}$ = 4.3 $\times$
10$^{-2}$ f$_{H\alpha}^{-1/2}$ cm$^{-3}$ and M$_{H\alpha}$ =8.0
$\times$ 10$^6$ f$_{H\alpha}^{1/2}$ M$_{\sun}$ respectively.  If we
make the physically reasonable assumption that this cooler gas is in
rough pressure balance with the hot X-ray gas, then the minimum
pressure of 1.3$\times$ 10$^{-11}$ dynes cm$^{-2}$ would require that
the H$\alpha$ emitting gas is highly clumped (f$_{H\alpha}$ $\sim$ 8
$\times$ 10$^{-5}$). This in turn implies that n$_{e,H\alpha}$ = 5
cm$^{-3}$ and M$_{H\alpha}$ =1.1 $\times$ 10$^5$ M$_{\sun}$. We note
that a small volume filling-factor for this gas is consistent with the
complex knotty, filamentary structure seen in the deep H$\alpha$ image
in DB.

The total bolometric luminosity and ionizing luminosity of M~82
(\cite{McL93} and references therein) can be used together with the
starburst models of \cite{LH95} to predict the rate at which the
starburst injects mechanical energy and momentum.  While these
predictions will depend to some degree on the evolutionary state of the
starburst (cf. Figures 66 through 69 in Leitherer \& Heckman) we
estimate that the mechanical luminosity and momentum flux provided by
M~82 are $\sim$ 2.5 $\times$ 10$^{42}$ erg s$^{-1}$ and $\sim$ 2
$\times$ 10$^{34}$ dynes respectively. For this wind mechanical
luminosity, the rate at which the X-ray ridge intercepts this energy is
2.5 $\times$ 10$^{41}$ $\Omega_{w}$$^{-1}$ erg s$^{-1}$, where
$\Omega_{w}$ is the total solid angle into which the wind flows
($\Omega_{w}$ $\leq$ 4$\pi$). Comparing this heating rate to the X-ray
luminosity of the ridge (2.2 $\times$ 10$^{38}$ erg s$^{-1}$) shows
that the wind could easily power the emission. It could also supply the
ridge's estimated thermal energy in a timescale of 10$^{6}$
f$_X$$^{1/2}$ $\Omega_{w}$ years (much less than the
wind/starburst lifetime).

Matching the observed high pressure in the X-ray ridge is more
challenging.  A wind momentum flux of 2.5 $\times$ 10$^{34}$ dynes
leads to a wind ram pressure at a radius of 11.6 kpc of 2 $\times$
10$^{-11}$ $\Omega_{w}$$^{-1}$ dyne cm$^{-2}$.  Consistency with the
X-ray pressure estimated above would require that the X-ray gas has a
large volume filling factor (f$_X$ $\sim$ unity) and that the M~82 wind
is rather well collimated ($\Omega_{w}$ $\sim$ 1.5 steradian).  This
latter constraint is consistent with the morphology of the outflow
(e.g., \cite{G90}; \cite{McK95}; \cite{SBH98}).  Alternatively, we can
use the empirical measurements of the gas pressure in the inner region
of the M~82 nebula (\cite{HAM90}) and extrapolate these outward,
assuming that the wind ram pressure falls as r$^{-2}$. The pressure at
a radius of 1 kpc was measured to be $\sim$ 5 $\times$ 10$^{-10}$ dyne
cm$^{-2}$, and so the predicted wind ram pressure at a radius of 11.6
kpc would be $\sim$ 4 $\times$ 10$^{-12}$ dyne cm$^{-2}$, about a
factor three smaller than the estimated value (for f$_X$ = 1). In view
of the uncertainties in the estimated physical parameters for the X-ray
ridge, we regard this level of disagreement as acceptable.

If the optical line emission is also shock-excited by the outflowing
wind, then the luminosity observed at H$\alpha$ should be consistent
with the wind energetics.  We can estimate the amount of emission from
a shock as follows.  The total energy dissipated in a shock is,
$L_{shock} = 1/2 \rho A v^3_s= 1.0 \times 10^{34} n_0 A_{kpc} v^3_s$
ergs s$^{-1}$, where $A_{kpc}$ is the shock surface area in kpc$^2$,
$n_0$ is the preshock density (cm$^{-3}$), and $v_s$ is in units of km
s$^{-1}$.  From \cite{BDT85}, we have using the
equation above, L$(H\alpha)_{shock} \approx 3 L(H\beta)_{shock} \approx
3 L_{shock}/100 = 1.7 \times 10^{41} n_0 A_{kpc} (v_s/820)^3$ ergs
s$^{-1}$, if 90 km s$^{-1}$ $<$ $v_s$ $<$ 1000 km s$^{-1}$.  Adopting
the geometry above, the total surface area that the ridge would present
to the wind would be 3.7 by 3.7 kpc.  The observed H$\alpha$ luminosity
of the ridge (2.4 $\times$ 10$^{38}$ erg s$^{-1}$) then requires that
$n_0 (v_s/820)^3$ = 1.0 $\times$ 10$^{-4}$, while the pressure
estimated in the X-ray ridge (1.3$\times$ 10$^{-11}$ dynes cm$^{-2}$)
requires $n_0 (v_s/820)^2$ = 8 $\times$ 10$^{-4}$. Thus, provided that
the shock speed is greater than 100 km s$^{-2}$, it can both produce the
observed H$\alpha$ emission-line luminosity and be consistent with the
pressure estimated for the hot shocked gas observed in the X-rays.

\subsection{The Relationship Between the H$\alpha$ and X-Ray Emission}

What is the physical relation between the H$\alpha$ and X-ray emission
in this type of model in which the superwind drives a shock into an
ambient cloud in the halo of M~82? One possibility is that the shock
driven into the cloud by the wind is fast enough (i.e., $\sim$ 800 km
s$^{-1}$) to produce the X-ray emission, and that the shocked gas then
cools radiatively down to temperatures of-order 10$^4$ K and produces
the observed H$\alpha$ emission via recombination.  The difficulty with
this simple model is that the estimated radiative cooling time in the
X-ray gas ($\sim$ 3 $\times$ 10$^8$ years) is very long compared to the
characteristic dynamical time for the X-ray ridge: t$_{dyn}$ $\sim$
(0.9 kpc/820 km s$^{-1}$) $\sim$ 1 $\times$ 10$^6$ years. Simply put,
this model would require the X-ray ridge to be $\sim$ 300 kpc thick and
the H$\alpha$ emission would be at the back edge of this very thick
cooling zone!  An additional persuasive argument against this model is
provided by the rather quiescent kinematics of the H$\alpha$-emitting
material reported by DB. Their optical spectra show that the H$\alpha$
emission in the ridge spans a velocity range of only about 10$^2$ km
s$^{-1}$, which is inconsistent with material cooling behind a shock
with a velocity nearly an order-of- magnitude larger. While the
relatively small difference DB report between the mean velocity of the
ridge and M~82 proper (blueshift of 50 to 200 km s$^{-1}$) could perhaps
be explained as a projection effect in an outflow seen nearly in the
plane of the sky, this effect could not explain the narrowness of the
H$\alpha$ emission-line in the ridge ($\sim$ 10$^2$ km s$^{-1}$)
without appealing to an unreasonably idealized shock geometry (which is
bellied by the morphological complexity of the H$\alpha$ emission in the
image published by DB).

A much more plausible alternative is that we are actually observing {\it
two} physically-related shocks in the M~82 ridge.  We propose that as
the superwind encounters a large cloud in the halo of M~82, it drives a
relatively slow (of-order 10$^2$ km s$^{-1}$) radiative shock into the
cloud.  At the same time, this encounter also results in a fast (820 km
s$^{-1}$) stand-off bow-shock in the superwind fluid upstream from the
cloud.  The slow cloud shock (aided and abetted by ionizing radiation
from M~82) produces the H$\alpha$ emission.  This is consistent with the
quiescent H$\alpha$ kinematics.  The fast bow-shock in the wind produces
the X-ray emission and is predicted to only be offset from front edge of
the cloud by a fraction of the size of the cloud (\cite{kmc94};
\cite{bw90}).  Pressure balance between the two shocks requires that
$\rho_{wind} v_{bow}^2 = \rho_{cloud} v_{shock,cloud}^2$, where
$\rho$$_{wind}$ and $\rho$$_{cloud}$ are the preshock densities in the
wind and cloud respectively.  The observed X-ray pressure and
temperature imply that v$_{bow}$ $\sim$ 820 km s$^{-1}$ and
n$_{wind,ion}$ $\sim$ 1 $\times$ 10$^{-3}$ cm$^{-3}$.  Thus - for
example - if the shock speed in the cloud is v$_{s,cloud}$ = 100 km
s$^{-1}$, then n$_{cloud} \sim$ 0.07 cm$^{-3}$.  We note further that
the velocity of the bow shock is just the difference between the outflow
velocity of the superwind itself and that of the shocked cloud.
Assuming that latter velocity is much smaller than the former (as is
suggested by the kinematic information provided by DB), the wind
velocity implied by the X-ray temperature is $\sim$ 820 km s$^{-1}$.  To
drive a wind with a terminal velocity of 820 km s$^{-1}$, hot gas would
need to be injected into the starburst at a mean mass-weighted
temperature of 1 $\times$ 10$^7$ K (\cite{CC85}).  This temperature is
consistent with X-ray spectroscopy of the hot gas in the core of M~82
(DWH, \cite{ML97}; \cite{SPS97}).

We have seen above that the radiative cooling time in the X-ray ridge is
$\sim$300 times longer than the characteristic dynamical time for the
ridge.  This implies that the wind bow shock is adiabatic.  This can
also be seen by comparing the total rate of energy dissipation in the
shock ($L_{shock} = 1/2 \rho_{wind} A_{ridge} v_{bow}^3 = 7 \times
10^{40}$ erg s$^{-1}$ for the parameters estimated above) to the
observed X-ray luminosity of the ridge (2.2 $\times$ 10$^{38}$ erg
s$^{-1}$).  Thus, the thickness of the X-ray ridge in the direction of
the outflow is set by the dimensions of the region where adiabatic
expansion and cooling causes the X-ray surface-brightness to drop
precipitously.  This region will be of-order the size of the obstacle in
the flow that has caused the bow-shock to develop (the cloud).  This is
consistent with the observed X-ray morphology and the good spatial
correlation seen between the H$\alpha$ and X-ray emission.  We have also
estimated previously that the superwind can supply the ridge with its
estimated thermal energy content in a timescale of-order 10$^6$ years.
This is roughly equal to the dynamical time in the ridge (the timescale
for adiabatic cooling), as required.

We conclude that it is energetically feasible for the cloud to be
shock-heated by a superwind driven by the starburst in M~82.  A
wind/cloud shock interpretation leads to no uncomfortable estimates of
the luminosity, covering fraction, or time scales.  In addition, as
noted previously, there is ample evidence that M~82 lies in an ``HI
halo'' (\cite{Y93}, 1994).  It is likely that the outflowing gas
has intercepted and shock-heated some of this halo material, which is
observed to be clumpy over large scales (see \cite{Y93}).  We also
note that perhaps one may have expected to find stronger evidence {\it
a priori} for such interactions on the north side of the plane of M~82
compared to the south side: \cite{Y94} find that the velocity
gradient in the HI is much higher on the northern side (12 km s$^{-1}$
kpc$^{-1}$) of the plane of M~82 compared to that on the southern side
(3--6 km s$^{-1}$ kpc$^{-1}$).  This higher gradient may be related to
our observation of co-spatial H$\alpha$ and thermal X-ray emission at
large distances from the plane of M~82.

\subsection{Where is the Cloud?}

We have shown that a simple model of the interaction between a fast
($\sim$ 800 km s$^{-1}$) wind and a massive cloud can quantitatively
explain the physical and dynamical properties of the ridge. However, it
a fair question to ask why the putative cloud is not observed in the
sensitive HI map published by \cite{Y93}. More specifically, we
have estimated above that the high pressures in the X-ray gas and the
low velocity of the shock driven into the cloud by the wind together
mean that the preshock density of the cloud must be of-order 10$^{-1}$
cm$^{-3}$. For a cloud with a line-of-sight dimension of 3.7 kpc, the
implied cloud column density is then $\sim$10$^{21}$ cm$^{-2}$. The HI
map in \cite{Y93} constrains the HI column density of the ridge
to be $<$ 2.7$\times$ 10$^{19}$ cm$^{-2}$. This implies that the cloud
must now be fully ionized.

This could be accomplished by either the wind-driven shock or by the
ionizing radiation from M~82. The time for the shock to traverse the
cloud will be $\sim$ 10$^{7}$ years for $v_{shock,cloud}$ = 100 km
s$^{-1}$, comparable to the estimated age of the starburst and its
superwind. Similarly, the velocity of a photo-ionization front moving
though the cloud is just $v_i \approx \Phi_{LyC} n_{cloud}^{-1}$ (e.g.,
\cite{O89}), where $\Phi_{LyC}$ is the incident flux of ionizing
radiation.  The observed H$\alpha$ luminosity and the assumed
cross-sectional area the cloud presents to the starburst (3.7 by 3.7
kpc) imply that the average value of $\Phi_{LyC}$ is 1.5 $\times 10^{6}
f_{abs}^{-1}$ s$^{-1}$ cm$^{-2}$, where $f_{abs}$ is the fraction of
the incident ionizing photons absorbed by the cloud (i.e., allowing the
cloud to be optically-thin).  For a preshock density of 0.07 cm$^{-3}$
in the cloud, the implied velocity is $v_i \approx$ 210
f$_{abs}$$^{-1}$ km s$^{-1}$, and the time for the front to cross the
cloud is about 4 $\times$ 10$^6$ f$_{abs}$$^{-1}$ years. Thus it is
plausible that an initially neutral cloud like those seen in the HI
maps of Yun et al. (1993,1994) has indeed been fully ionized in the time
since the starburst and its superwind turned on.

Our rough estimates suggest that the timescale for the cloud to be
fully photoionized is likely to be somewhat shorter than the timescale
for it to be fully shock-ionized.  If correct, this would mean that the
H$\alpha$ emission produced by the gas cooling and recombining behind
the shock would have a high density and pressure, while the rest of the
ionized cloud (the as-yet unshocked material) would be at a much lower
pressure and density. We speculate that the bright knots and loops of
H$\alpha$ visible in the image in DB may correspond to shock-excited
emission from gas with a high pressure, while the fainter and more
diffuse emission may come from the low-pressure starburst-photoionized
gas that has not yet been shocked.

\subsection{Implications for the ``Superwind'' Phenomenon}

It is often difficult to associate discrete X-rays sources seen in the
halos of starburst galaxies with activity in the starburst galaxy
itself.  Often, it is assumed that most discrete objects observed in
the halos of starbursts are more distant objects (usually QSOs) seen
in projection.  Finding a spatial association between H$\alpha$
emission at the redshift of the starburst galaxy and X-ray emission is
a irrefutable way of being able to associate the X-ray emission
directly with activity within the starburst galaxy.

Having argued that this co-spatial region of soft X-ray and H$\alpha$
emission must be due to the outflowing superwind, it is now evident
that at least some starburst galaxies are able to drive material to
much larger distances from their nuclei than would be inferred using
the relatively high surface brightness X-ray or line emission that is
generally observed along their minor axes (see e.g., \cite{LH96};
\cite{HAM90}; DWH).  In M~82 for example, the inner ``continuous''
region of bright H$\alpha$ and X-ray emission extends only about 5
arcminutes from the nucleus along the minor axis.  The detached
``ridge'' of emission that we have discussed is at a projected angular
separation from the nucleus of over twice this distance (about 11
arcminutes).  Thus without making the association between the H$\alpha$
and X-ray emission, a variety of interpretations for the source of the
X-ray emission from this region would be possible.  This obviously has
important implications for our understanding of superwinds.  It
emphasizes that one must be careful in using the observed spatial
scales of the extended emission in starbursts to either access whether
or not the wind material is able to escape the potential of the host
galaxy or to estimate the dynamical age of the outflow.

It is interesting in this vein to consider the long-term fate of the
hot gas in the X-ray ridge of M~82. To do so, we will compare the
observed temperature of the gas to the ``escape temperature'' at that
location in the M~82 halo. Following \cite{W95}, the escape temperature
for hot gas in a galaxy potential with an escape velocity v$_{es}$ is
given by: T$_{es}$ $\simeq$ $1.1\times 10^{5}$ (v$_{es}$/100 km
s$^{-1}$)$^{2}$ K.  The combined CO plus HI rotation curve of M~82 shows
a roughly Keplerian decrease with radius for radii greater than 0.4
kpc, and $v_{rot}$ has dropped to 60 km s$^{-1}$ at a radius of 3.5 kpc
(\cite{S98}). If this Keplerian fall-off continues to larger radii, the
implied escape velocity at a radius of 11.6 kpc is only 45 km s$^{-1}$
and the corresponding escape temperature is T$_{es}$ $\simeq$ $2\times
10^{4}$ K. This is 450 times cooler than the observed X-ray emission.
Even if we assume instead that M~82 has an isothermal dark matter halo
with a depth corresponding to a circular velocity of 60 km s$^{-1}$
(the maximum permitted by the directly measured rotation curve) and
further assume that this halo extends to a radius of 100 kpc, the
escape velocity at r = 11.6 kpc is about 150 km s$^{-1}$ and T$_{es}$ =
2.5 $\times 10^5$ K. This is still a factor of 36 below the observed
temperature of the gas in the ridge.  This implies that the observed
hot gas (and by inference, the superwind itself) will be able to escape
the gravitational potential of M~82.

The X-ray/H$\alpha$ ridge has particularly interesting implications for
the physics of superwind emission. While the existence of superwinds is
now well-established, the process(es) by which they produce X-ray
emission is (are) not clear. This topic has been nicely reviewed by
\cite{Str98}.  If the outflowing wind consists purely of the
thermalized ejecta from supernovae and stellar winds, the wind fluid is
hot ($\sim$ 10$^8$ K prior to adiabatic cooling) and is so tenuous that
it will be an insignificant source of X-ray emission (\cite{CC85};
\cite{Suc94}). Various alternatives have been proposed to boost the
X-ray luminosity produced by a superwind: 1) Substantial quantities of
material could be mixed into the wind in or near the starburst as the
stellar ejecta interact with the starburst ISM (the wind could be
centrally ``mass-loaded'' - \cite{Suc96}; \cite{H97}). 2) Large
quantities of gas could be added {\it in situ} as the wind overtakes
and then evaporates or shreds gas clouds in the galactic halo (e.g.,
\cite{Suc94}). These clouds could either be pre-existing clouds in the
halo or material from the disk of the starburst galaxy that has been
carried into the halo by the wind.  3) The tenuous wind fluid could
drive a shock into a denser volume-filling galactic halo, with the
observed X-ray emission arising from the shocked-halo material rather
than the wind-fluid itself.

In the case of the ridge of emission in the halo of M~82, we have argued
that the observable X-ray emission arises as the wind fluid -- which
due to adiabatic expansion and cooling would otherwise have an X-ray
surface-brightness below the level detectable in the existing ROSAT
image -- encounters a few-kpc-size-scale cloud in the halo. We have
argued that the observed X-ray emission is produced in the bow-shock
created in the superwind just upstream of the cloud. This process might
be particularly relevant to the case of M~82, which is immersed in an
extensive system of tidally-liberated gas clouds, shared with its
interaction partners M81 and NGC3077.  Such material would also provide
a potential source of optical line emission by either intercepting a
small portion of the ionizing radiation produced by the central
starburst or as it is shock-heated by the superwind. Given the strong
connection between galaxy interactions/mergers and the starburst
phenomenon (e.g., \cite{SM96} and references therein), this
mechanism may be generally applicable to superwinds.

\section{Summary and Conclusions}

We report an analysis of a newly-discovered (\cite{DB99})
region of spatially-coincident X-ray and H$\alpha$ emission to the
north of the prototypical starburst/superwind galaxy M~82.  At the
distance of M~82 (3.63 Mpc) the dimensions of this correlated ridge of
emission are 3.7 $\times$ 0.9 kpc and it lies at a projected distance
of about 11.6 kpc above the plane directly along the minor axis of M~82
(and hence along the axis of the superwind).  We measure a total
H$\alpha$ flux from the ridge of 1.5 $\times$ 10$^{-13}$ ergs s$^{-1}$
cm $^{-2}$.  This flux yields a total H$\alpha$ luminosity of 2.4
$\times$ 10$^{38}$ ergs s$^{-1}$. The total H$\alpha$ flux for the
whole of M~82 is approximately 4.6 $\times$ 10$^{-11}$ ergs s$^{-1}$ cm
$^{-2}$.  Thus the H$\alpha$ ridge constitutes approximately 0.3\% of
the total observed H$\alpha$ flux from the galaxy. X-ray emission is
seen over the same region as the H$\alpha$ emission by the ROSAT PSPC
and HRI. With the 15 arcsec (260 pc) off-axis spatial resolution of the
HRI, the ridge is diffuse and featureless except for a (probably
unrelated) soft point source at its ENE end.  The best fit to the X-ray
emission from the ridge is a single thermal component with a
temperature of kT=0.80$\pm$0.17 keV, absorbed by a column density N$_H$
= 3.7 $\times$ 10$^{20}$ cm$^{-2}$.  The absorption is consistent with
the foreground Galactic column.  The total unabsorbed flux from the
ridge of emission is 1.4 $\times$ 10$^{-13}$ ergs cm$^{-2}$ s$^{-1}$
and its luminosity of 2.2 $\times$ 10$^{38}$ erg s$^{-1}$ comprises
about 0.7\% of the total X-ray emission from M~82 in the 0.l -- 2.4 keV
ROSAT band.

We find that the observed H$\alpha$ emission from the ridge could be
produced via photoionization by the dilute radiation from the starburst
propagating into the M~82 halo along the path cleared by the superwind.
On the other hand, the X-ray emission from the ridge can be most
naturally explained as the result of shock-heating associated with an
encounter between the starburst-driven galactic ``superwind'' and a
large photoionized cloud in the halo of M~82. Shock-heating may also
contribute significantly to the excitation of the H$\alpha$ emission,
especially the regions of the highest surface brightness. We find that
the H$\alpha$ and X-ray luminosities and the X-ray temperature and
estimated gas pressure can all be quantitatively understood in this
context.  The specific model we favor is one in which the encounter
between the superwind and the cloud drives a slow radiative shock into
the cloud (producing some of the H$\alpha$ emission) and also leads to
an adiabatic X-ray-emitting bow-shock in the superwind fluid upstream
from the cloud. The cloud is now fully ionized, and so is not detected
in the HI maps of Yun et al. (1993, 1994).  The wind-cloud interaction
process may be especially relevant to M~82, whose halo contains
tidally-liberated gas clouds. However, it may well occur generally in
powerful starbursts (which are usually triggered by galaxy interactions
or mergers).

The existence of a region of spatially coincident X-ray and H$\alpha$
emission with strong evidence for wind/cloud interaction is important
for several reasons. Firstly, our analysis sheds new light on the
physical processes by which galactic winds produce X-ray emission, and
implies that wind/cloud collisions are an important emission-producing
process. The emission ridge in M~82 highlights the potential pitfalls in
using the observed X-ray or H$\alpha$ emission to trace the size and
infer the ages or fates of galactic winds:  the winds may extend to
radii far larger than the region over which they have densities and
temperatures high enough to produce observable emission.  Wind-cloud
encounters then offer us the possibility of `lighting up' the wind at
radii that are otherwise inaccessible.  The X-ray ridge in M~82 implies
that the wind is able to reach large heights above the plane of the
galaxy, thereby increasing the likelihood that it will be able to
eventually escape the galaxy potential well. Indeed, the observed gas
temperature exceeds the local ``escape temperature'' from the M~82
gravitational potential well by about two orders-of-magnitude. This
result strengthens the case that starburst-driven galactic winds are
important sources of the metal-enrichment and heating of the
Intergalactic Medium.

\acknowledgements
The authors wish to thank Kitt Peak National Observatory for their
generous allocation of observing time and their staff for making sure
that the observations were carried out efficiently and effectively.  We
thank Ed Moran, David Strickland, and Crystal Martin for illuminating
discussions about the physics of superwinds, and David Devine for
communicating his discovery paper in advance of of its publication.  We
would like to express our sincerest thank you to the referee for
carefully reading the manuscript and for providing us with detailed
comments that have substantially improved this presentation.  This work
was supported in part by NASA LTSA grants NAGW-3138 to TH and NAGW-4025
to KW. This research has made use of the IRAF package provided by NOAO
and NASA/IPAC extragalactic database (NED), which is operated by the
Jet Propulsion Laboratory, Caltech, under contract with the National
Aeronautics and Space Administration.

\newpage

\centerline{FIGURE CAPTIONS}

\figcaption [] {The greyscale is the H$\alpha$ image and the contours
are the ROSAT PSPC image of M~82.  The lowest surface brightness
emission visible in the  H$\alpha$ corresponds to a flux of 3.5
$\times$ 10$^{-17}$ ergs s$^{-1}$ cm $^{-2}$ arcsec$^{-2}$.  The
spatial coincidence of the H$\alpha$ and X-ray emission is quite good.
The images were aligned using using the alignment determined for the
continuum image.  The coordinates are shown for J2000.}

\figcaption [] {The greyscale is a narrow-band continuum image (central
wavelength approximate 6658\AA) image and the contours are the smoothed
($\sigma$=10'' gaussian) ROSAT PSPC image of M~82.  The images were
aligned using what is likely to be a background AGN and galaxy that
were both visible in the optical continuum and X-ray images.  The
coordinates are shown for J2000.}

\figcaption [] {The Rosat PSPC spectrum of the diffuse X-ray emission
from the ridge to the right of the point source (circles) and
superimposed on the point source (crosses).  Data below 0.6 keV are
ignored for the latter case to eliminate the point source.  Also shown
are the best-fit MEKAL plasma models (see text) folded through the
instrumental response.}

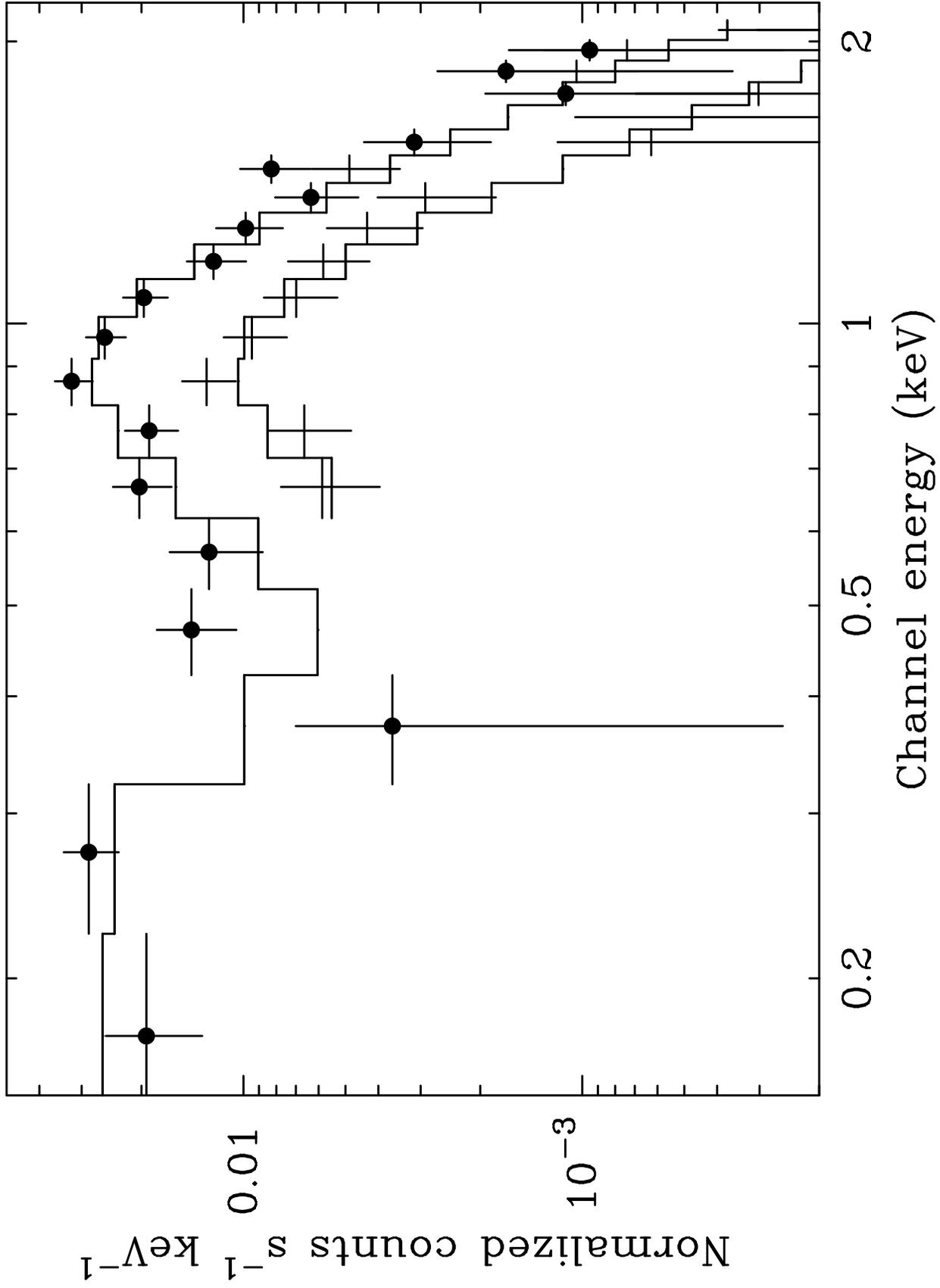
\end{document}